\documentclass{PoS}

\usepackage{graphicx}

\title{Gamma-rays in the radio galaxy 3C 84: A complex situation}

\ShortTitle{Gamma-rays in the radio galaxy 3C 84}

\author{Jeffrey A. Hodgson\\
        Korea Astronomy and Space Science Institute, Daejeon, South Korea\\
        E-mail: \email{jhodgson@kasi.re.kr}}

\author{\speaker{Bindu Rani}\\
        NASA Goddard Space Flight Center, Greenbelt, MD, 20771, USA\\
        E-mail: \email{bindu.rani@nasa.gov}}

\author{Junghwan Oh\\
        Seoul National University, Gwanak-gu, Seoul, South Korea\\
        E-mail: \email{joh@snu.ac.kr}}

\abstract{3C\,84 is a nearby Active Galactic Nucleus (AGN) that is unique in that is believed that we are observing near the true jet launching region - unlike blazars. The source is active in $\gamma$-rays and has been detected with \emph{Fermi} since its launch in 2008, including being detected at TeV energies with other instruments. Due to the relative proximity of the source (z=0.018), it provides a unique opportunity to pinpoint the location of the $\gamma$-ray emission by combining the $\gamma$-ray data with very long baseline inteferometry (VLBI) data. A study using the Korean VLBI network (KVN) showed that the $\gamma$-rays occur in both downstream jet emission and the region near where the jet is launched. Further analysis of the kinematics using Wavelet Image Segmentation and Evaluation (WISE) algorithm, which uses 2-dimensional cross-correlations to statistically derive the kinematics of high-resolution 7\,mm VLBA data show that the $\gamma$-ray emission is caused by a fast-travelling shock catching a slower moving shock and then interacting with the external medium, in behaviour reminiscent of a long duration gamma-ray burst (GRB). This could explain why such high energy flaring is seen in such low Doppler boosted sources. Finally, we show some early results from a study of the jet launching region using the Global mm-VLBI Array (GMVA). The nucleus appears to have a consistent double nuclear structure that is likely too broad to be the true jet base. }

\FullConference{7th Fermi Symposium 2017\\
		15-20 October 2017\\
		Garmisch-Partenkirchen, Germany}

\begin{document}

\section{Introduction}

3C\,84 (the radio counterpart of NGC\,1275) is a mis-aligned active galactic nucleus (AGN) located at the center of the Perseus cluster. As one of the brightest radio sources, it has a long history of study and has been known to be bright at radio wavelengths since at least the 1950s \cite{baade_win_54,BBS63,BB65}. The source is known to be two galaxies colliding and known to have large inhomogenous free-free absorption in the central region \cite{haschick82,walker2000,fujita17}. \\

\begin{figure*}\label{7mm}\centering
\includegraphics[width=0.5\textwidth]{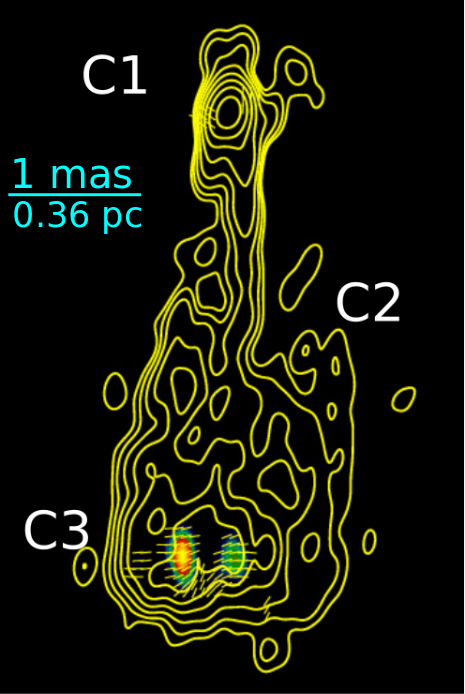}
\caption{\textbf{A recent (March 2017) map of 3C\,84 adapted from public data available from the BU-VLBA-BLAZAR program, see \cite{jor17} for details. The contours are set at The broad emission regions C1, C2 and C3 are annotated. Colours denote polarised intensity. Contours decrease by a factor of $\sqrt{2}$ from the peak flux density, with 1\,mas corresponding to 0.359\,pc. } }
\end{figure*}

The current radio morphology of the sub-parsec scale structure is shown in Fig. \ref{7mm}. The C1 region is thought to be at or near the location of the jet base, C3 is a slow moving feature and C2 is a quasi-stationary or very slow moving feature. While the source has been observed with Very Long Baseline Interferometry (VLBI) since the 1980s \cite{romney84,backer87,marr89,backer87,krichbaum92}, its radio brightness had been decreasing. However, since $\sim$2005, the source has been rising in radio flux densities, apparently related to the ejection of a new component known as C3 from near the central super-massive black hole (SMBH) \cite{nagai10}. More recently the source has been detected in $\gamma$-rays and also at higher frequencies, possibly related to the interaction of the C3 region with the external medium \cite{nagai14,nagai17,3c84_atel16_1, 3c84_atel16_2, 3c84_atel16_3, 3c84_atel16_4, 3c84_atel16_5}. The nature of the radio emission and its connection with the $\gamma$-ray emission is an area of active study. In this proceeding, we briefly summarise the results of one submitted paper and two additional papers examining the $\gamma$-ray emission, high-resolution kinematics and morphology of 3C\,84. \\

We have used a flat $\Lambda$CDM cosmology with $H_{0}$=69.6 and $\Omega_{m}$=0.286 \cite{bennett14}, which corresponds to a linear scale of 1\,mas = 0.359\,pc at a luminosity distance of $D_{L}$=74.047\,Mpc. \\

% \begin{figure*}\label{gmva}\centering
% \includegraphics[width=0.7\textwidth]{gmva_may14.png}
% \caption{\textbf{Example GMVA map of 3C\,84 in May 25 2014. Note the double nuclear structure.} }
% \end{figure*}

\section{Radio-Gamma correlations}

\begin{figure*}\label{LCs}
\includegraphics[width=\textwidth]{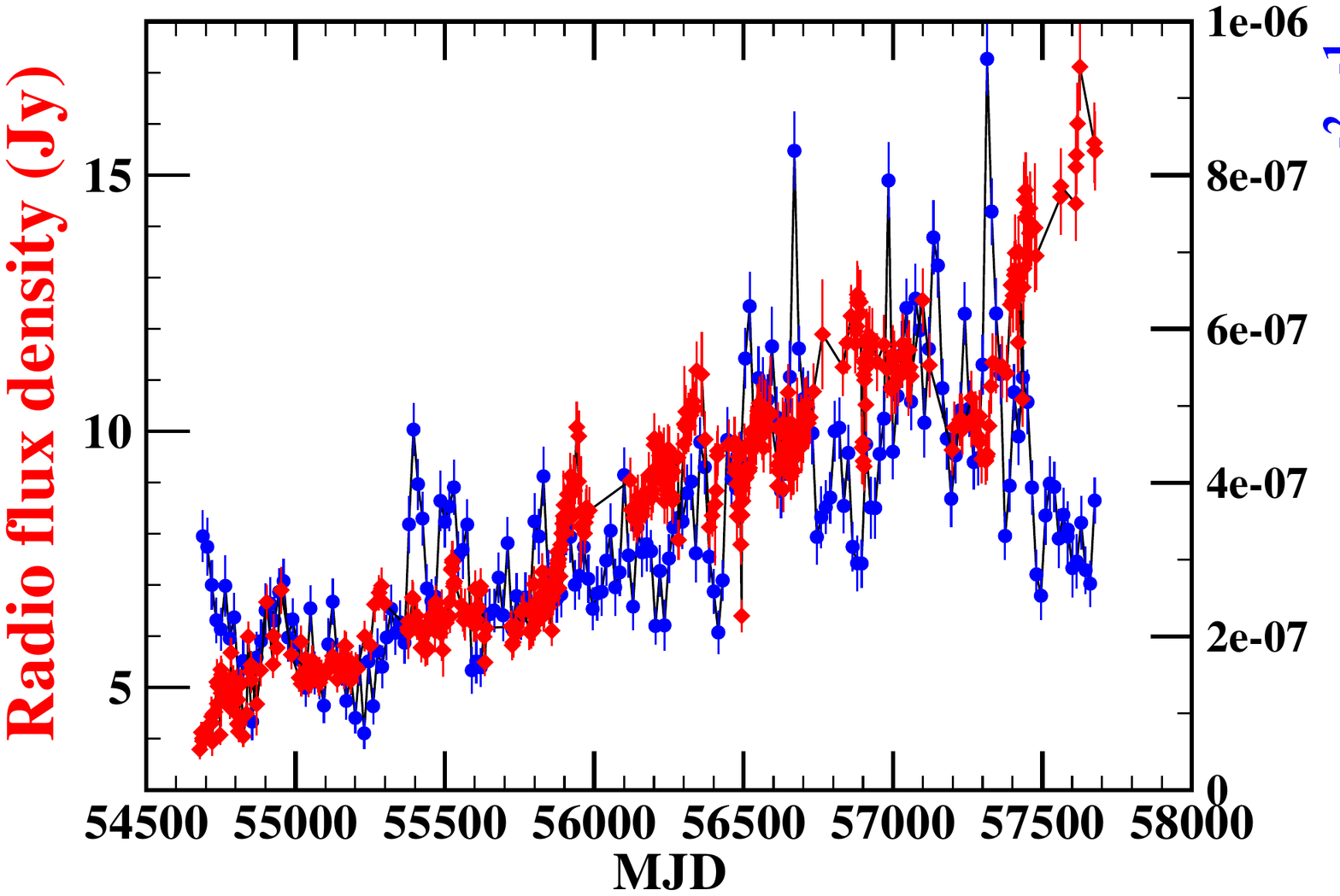}
\caption{\textbf{The total intensity light-curves of \emph{Fermi}-LAT $\gamma$-ray (blue) and 1\,mm data from the sub millimiter array (SMA) calibrator page (red) in 3C\,84. The units of the Gamma-ray photon flux are in ph\,cm$^{2}$ \,s$^{-1}$. } }
\end{figure*}

\begin{figure*}\label{dcf}
\includegraphics[width=\textwidth]{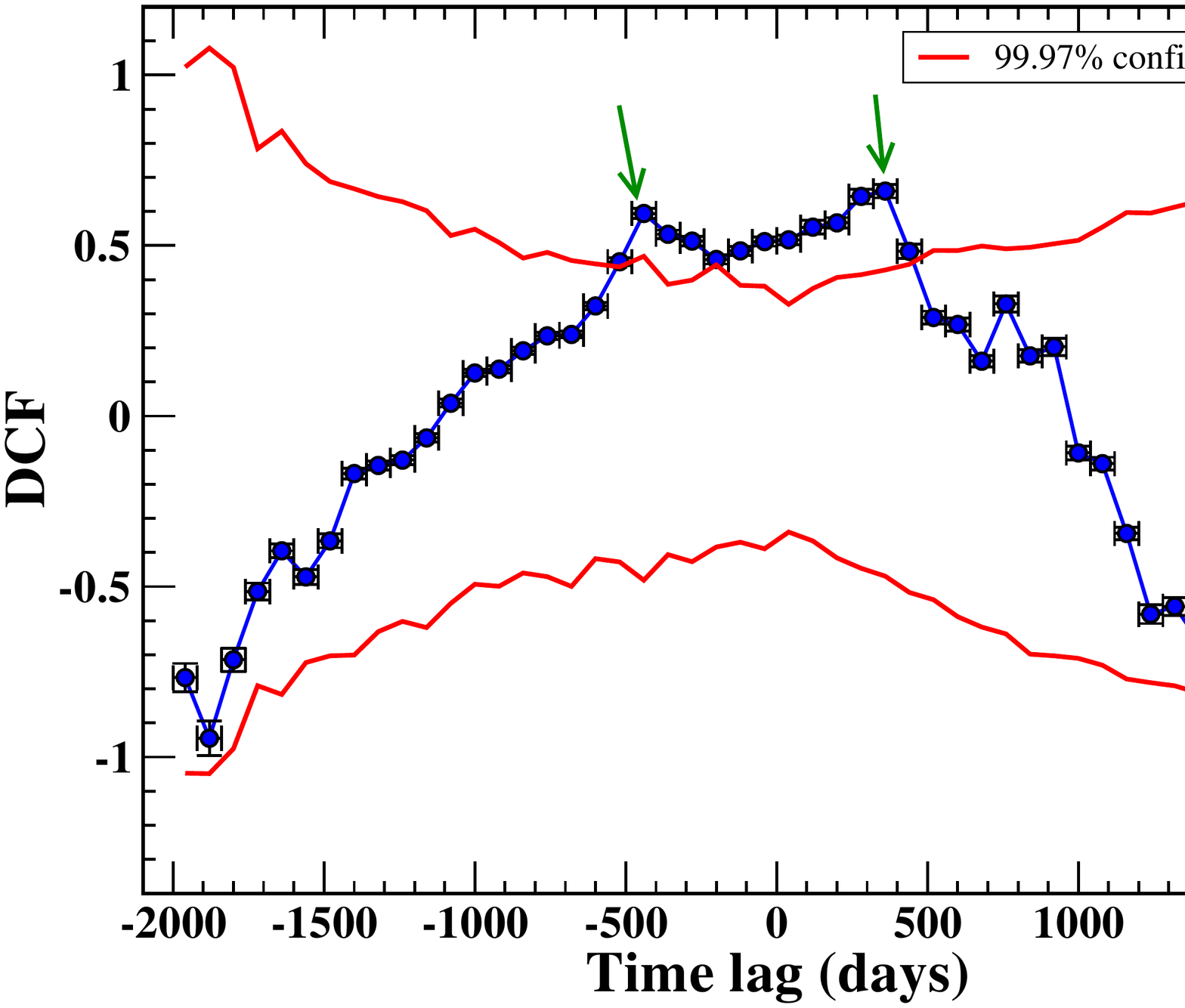}
\caption{\textbf{DCF of monthly binned $\gamma$-ray data with 1\,mm total intensity radio light-curves.} }
\end{figure*}

Here we present some highlights from the upcoming paper (Hodgson et al. MNRAS, submitted). Variations in $\gamma$-ray emission from the \emph{Fermi} LAT $\gamma$-ray telescope are compared with total intensity 1\,mm data and monthly multi-frequency observations (at 14\,mm, 7\,mm, 3\,mm and 2\,mm) using the Korean VLBI Network (KVN). In this paper, our results show that there are large flares (particularly since 2015), short-time scale variability and a slow rising trend (see Fig. \ref{LCs}). The slow rising trend and large flares were located in the slow-moving C3 region. A discrete correlation analysis (DCF) of the $\gamma$-ray and total intensity 1\,mm light-curves (see Fig. \ref{dcf}) found that they were highly correlated with an approximate 8 month lag. The data were interpolated and regridded to be analysed using a Pearson correlation coefficient. The large $\gamma$-ray flaring since 2015 was highly significantly correlated with the C3 region, however before 2015 the short time-scale variations were significantly correlated with the C1 region, showing that $\gamma$-rays could be associated both with the region near the SMBH (C1) and a travelling shock (C3). 

\section{Kinematic Analysis}

In a further upcoming paper (Hodgson et al. in prep), the kinematics of the source was analysed using the Wavelet Image Segmentation and Evaluation (WISE) analysis package, which had previously been used to analyse the kinematics in M87 \cite{mertlov15,mertlov16,mertlovm87}. This analysis revealed a highly complex kinematic structure. In a broad sense, non-ballistic component motions were detected, consistent with a sheath of a helical jet. The motion of C3, takes the appearance of ``crashing through'' the C2 region, which is itself revealed to be a very slow-moving jet component ejected in the early 1980s. \\

The WISE analysis shows that apparent super-luminal motion ($\sim$1.2\,c) are detected in the outer sheath of the jet, along the same trajectory as earlier slower moving shocks. When these faster shocks interact with slower moving shocks, $\gamma$-ray emission is seen. When these shocks then hit the external medium, even larger $\gamma$-ray flaring is seen. This behaviour appears to be reminiscent of the behaviour expected in a long-duration $\gamma$-ray burst (GRB). This could perhaps be an explanation for the so-called ``Doppler crisis'' where less Doppler-boosted sources are often seen at TeV energies, whereas Doppler boosted ``blazar'' sources are less frequently detected \cite{jor17}. 

\section{GMVA Analysis}

We have additionally (Oh and Hodgson et al. in prep) analysed six epochs of 3C\,84 using the Global mm-VLBI Array (GMVA, see \cite{hodgson16b} for details). The source was observed between 2008 and 2016. These observations achieve angular resolution of greater than 50 micro-arcseconds, corresponding to a linear resolution on the order of $\sim$hundreds of $R_{S}$. In all epochs since 2008, we observe a double nuclear structure (C1a and C1b) separated by $\sim$0.7\,mas, with a trend of increasing brightness temperature in the north-easterly and south-westerly directions. We find that the width of the 3\,mm VLBI core is likely to broad to be the true base of the jet, or directly related to the accretion disk. \\

\section{Conclusion}

We have shown that the $\gamma$-rays in 3C\,84 occur in at least two different locations, including in the jet launching region and in travelling knots. Superluminal motion is for the first time detected in 3C\,84 and is associated with $\gamma$-ray flaring when the fast shock ``catches up'' with a slower shock. Larger flaring including at TeV is then seen when these shocks interact with the external medium. This behaviour is similar to what is expected in long-duration GRBs, suggesting a possible solution to the ``Doppler crisis''. Additionally, the GMVA data suggest that the 3\,mm VLBI core is likely not the true base of the jet.

\section{Acknowledgements}

The $Fermi$-LAT Collaboration acknowledges support from a number of agencies and institutes for both development and the operation of the LAT as well as scientific data analysis. These include NASA and DOE in the United States, CEA/Irfu and IN2P3/CNRS in France, ASI and INFN in Italy, MEXT, KEK, and JAXA in Japan, and the K.~A.~Wallenberg Foundation, the Swedish Research Council and the National Space Board in Sweden. Additional support from INAF in Italy and CNES in France for science analysis during the operations phase is also gratefully acknowledged. The research at Boston University was supported by NASA through a number of Fermi Guest Investigator program grants, most recently NNX14AQ58G. The St. Petersburg University group acknowledges support from the Russian Science Foundation grant 17-12-01029. This research is partially based on observations performed at the 100\,m Effelsberg Radio Telescope, the IRAM Plateau de Bure Millimetre Interferometer, the IRAM 30\,m Millimeter Telescope, the Onsala 20\,m Radio Telescope, the Mets\"{a}hovi 14\,m Radio Telescope, the Yebes 30\,m Radio Telescope and the Very Long Baseline Array (VLBA). The VLBA is an instrument of the Long Baseline Observatory. The Long Baseline Observatory is a facility of the National Science Foundation operated by Associated Universities, Inc. Single-dish data were acquired through the FGAMMA program of the MPIfR and the Submillimeter Array (SMA) flux monitoring programs. The Submillimeter Array is a joint project between the Smithsonian Astrophysical Observatory and the Academia Sinica Institute of Astronomy and Astrophysics and is funded by the Smithsonian Institution and the Academia Sinica. This research has made use of the NASA/IPAC Extragalactic Database (NED) which is operated by the Jet Propulsion Laboratory, California Institute of Technology, under contract with the National Aeronautics and Space Administration. This research made use of Astropy, a community-developed core Python package for Astronomy \cite{astropy}.

\bibliographystyle{JHEP}
\bibliography{Bibliography} % if your bibtex file is called example.bib

\end{document}